\documentclass[amsmath,amssymb,superscriptaddress,twocolumn]{revtex4-2}
\usepackage{graphicx}% Include figure files
\usepackage{dcolumn}% Align table columns on decimal point
\usepackage{bm}% bold math
\usepackage[colorlinks=true, citecolor=blue, urlcolor = blue, linkcolor= red, bookmarks=true]{hyperref}
\begin{document}
\title{Thermodynamic stability and $P$--$V$ criticality of nonsingular-AdS black holes endowed with clouds of strings}
\author{Ashima Sood}
 \email{ashimasood1993@gmail.com}
\affiliation{Centre for Theoretical Physics, Jamia Millia Islamia, New Delhi 110025, India}
\affiliation{Department of Mathematics, Netaji Subhas University of Technology, New Delhi-110078, India}

\author{Arun Kumar}
\email{arunbidhan@gmail.com }
\affiliation{Centre for Theoretical Physics, Jamia Millia Islamia, New Delhi 110025,
		India}
\author{J. K. Singh}
\email{jksingh@nsut.ac.in}
\affiliation{Department of Mathematics, Netaji Subhas University of Technology,  New Delhi-110078, India}
	
\author{Sushant G. Ghosh} \email{sghosh2@jmi.ac.in }
\affiliation{Centre for Theoretical Physics, Jamia Millia Islamia, New Delhi 110025,
		India}
\affiliation{Department of Mathematics, Netaji Subhas University of Technology,  New Delhi-110078, India}
\affiliation{Astrophysics and Cosmology Research Unit,
School of Mathematics, Statistics and Computer Science,
University of KwaZulu-Natal, Private Bag X54001,
Durban 4000, South Africa}

\begin{abstract}
We investigate the extended phase space thermodynamics of nonsingular-AdS black holes 
minimally coupled to clouds of strings in which we consider the cosmological constant ($\Lambda$) as the pressure ($P$) of the black holes and its conjugate variable thermodynamical volume ($V$) of the black holes.  Owing to the background clouds of strings parameter ($a$), we analyse the Hawking temperature, entropy and specific heat on horizon radius for fixed-parameter $k$.  We find that the strings clouds background does not alter small/large black hole (SBH/LBH) phase transition but occurs at a larger horizon radius, and two second-order phase transitions occur at a smaller horizon radius.  Indeed, the $G$--$T$ plots exhibit a swallowtail below the critical pressure, implying that the first-order phase transition is analogous to the liquid-gas phase transition at a lower temperature and lower critical pressure. To further examine the analogy between nonsingular-AdS black holes and a liquid-gas system, we derive the exact critical points and probe the effects of a cloud of strings on $P-V$ criticality to find that the isotherms undergo liquid-gas like phase transition for $\tilde{T}\,<\,\tilde{T}_c$ at lower $\tilde{T}_c$.   We have also calculated the critical exponents identical with Van der Walls fluid, i.e.,  same as those obtained before for arbitrary other AdS black holes, which implies that the background clouds of strings do not change the critical exponents. 
\end{abstract}
\maketitle

\section{Introduction} 
The black hole's thermodynamics investigation has communicated a deep and fundamental relationship among gravitation, thermodynamics, and quantum theory. Interestingly, classical and semiclassical analysis has given rise to most of our present physical understandings of the nature of quantum phenomena \cite{Wald:1999vt,Israel:1967wq}.  An interesting phenomenon in the AdS black hole thermodynamics is the existence of the transition between AdS--Schwarzschild black hole and thermal AdS, unlocked in the pioneering work of Hawking and Page \cite{Hawking:1982dh}, which has attracted numerous astrophysicists towards studying black holes thermodynamics in AdS spacetimes. Indeed,  the Hawking-Page phase transition between large black holes (LBH) and thermal gas in the AdS space is a topic of intense research \cite{Dutta:2013dca,Li:2014ixn,Cai:2013qga,Cai:2014znn,Zhao:2013oza,Mo:2016ndm,Mo:2013sxa,Altamirano:2013uqa,Mo:2013ela,Altamirano:2013ane,Zou:2013owa,Ghaffarnejad:2018exz,Liu:2016uyd,Hansen:2016ayo,Ghosh:2020ijh}, e.g., an analogy between phase structures of various AdS black holes and statistical models associated with Van der Waals like phase transitions has been suggested \cite{Liu,Dolan:2010,Kubiznak:2012wp}. The analysis of AdS black hole phase transition has been generalized to the extended phase space where the cosmological constant has been treated as the pressure of the black hole revealing several exciting results\cite{Kastor:2009wy,Cvetic:2001bk,Belhaj:2013cva,He:2016fiz}.

In an interesting paper, the authors \cite{Herscovich:2010vr} explicitly bring out how the effect of a background of clouds of strings can modify black hole solutions and their properties aspects in non-extended thermodynamics. A cascade of subsequent interesting work analysed black hole solutions in clouds of strings model \cite{Ghosh:2014pga,Lee:2014dha,MoraisGraca:2018ofn,Lee:2015xlp,Singh:2020nwo} and these black holes have been shown to also exhibit Hawking-Page type phase transitions \cite{Ghosh:2014pga}.  The theoretical developments signal toward a scenario in
which the fundamental building blocks of the Universe are extended
objects instead of point objects can be exciting and have been taken quite
seriously \cite{synge} with the most natural candidate being
a one-dimensional strings object.  In addition, the intense level of activity in string theory has led to the idea that the static Schwarzschild black hole (point mass) may have
atmospheres composed of fluid or field of strings. The clouds of strings background for the  Schwarzschild black hole was proposed by Letelier \cite{Letelier:1979ej}  to demonstrate that event horizon has modified radius $ r_H={2M}/{(1-a)}$ with $0<a<1$ being the string cloud parameter \cite{Letelier:1979ej}, thereby enlarging the Schwarzschild radius of the black hole by the factor $(1-a)^{-1}$.  Many authors generalized the pioneering work of Letelier \cite{Letelier:1979ej}, for instance, in GR \cite{MoraisGraca:2018ofn},  for EGB models \cite{Ghosh:2014dqa, Ghosh:2014pga,Singh:2020nwo} and in Lovelock gravity \cite{Lee:2014dha,Lee:2015xlp}.  Thus, the study of black holes surrounded by clouds of strings, in  general
relativity or modified theories, may be critical because
relativistic strings at a classical level can be used to construct
applicable models \cite{synge}.  

The primary purpose of this paper is to look for an exact spherically symmetric nonsingular black hole solution endowed with clouds of strings in AdS spacetime, and explicitly mention how background clouds of strings can alter black hole solutions and their extended phase thermodynamic qualitative features we know from our experience in general relativity. 
We shall, in turn, probe the effects of background  on $P - V$ criticality, the critical behaviour of the thermodynamic quantities and demonstrate that there exists a phase transition and critical phenomena similar to the ones in a Van der Waals liquid-gas system. Thus, it is our goal to connect nonsingular-AdS black holes with the concept of black hole chemistry \cite{Kubiznak:2014zwa,Kumar:2020cve,Tzikas:2018cvs}. The black hole chemistry -- a new perspective on black hole thermodynamics, with the interpretation of black hole mass as enthalpy, the cosmological constant $\Lambda$ and  it's conjugate variable, respectively, as a pressure term and the thermodynamical volume of the black holes has led to a new understanding of Van der Waals fluids phase transitions from a gravitational viewpoint \cite{Kubiznak:2014zwa}. In particular, we investigate the analogy between nonsingular-AdS black holes endowed with clouds of strings and a liquid-gas system and find exact solutions of the critical points. 
 
This paper is organized as follows. In Section~\ref{Sec2}, we consider nonsingular--AdS black holes surrounded by clouds of strings, discuss the effect of clouds of strings parameter $a$, nonlinear electrodynamics (NED) parameter $k$, on horizon structure. The thermodynamic properties of the solution derived and discussed in Sec.~\ref{Sec3}.  
In Sec. \ref{Sec4}, the thermodynamical stability analysis and $P-V$ criticality of the black holes has been discussed, and the critical exponents have been calculated in Sec. \ref{Sec5}. Finally, we summarize our main findings in Section~\ref{conc}.  

We use geometrized units $8 \pi G = c= 1$, unless units are specifically defined.

\section{Nonsingular black holes with clouds of strings}\label{Sec2}
The AdS black holes are natural tools to investigate AdS/CFT correspondence and black hole thermodynamics in extended phase space. Here, we are interested in a nonsingular-AdS black hole solution endowed with clouds of strings, and the energy-momentum tensor of clouds of strings resembles the global monopole \cite{Singh:2020nwo}. The nonsingular-AdS black holes can be derived from general relativity minimally coupled to NED whose action reads
\begin{equation}\label{action}
\mathcal{S} = \frac{1}{16 \pi} \int \mathrm{d}^4x \sqrt{-g}  [ \mathcal{R}+6l^{-2} - \mathcal{L(F)} ] + {S_M},
\end{equation}
where $\mathcal{R}$ is the scalar curvature, $g$ is the determinant of the metric tensor and $l$ is positive AdS radius related to cosmological constant $\Lambda $ through the relation $ \Lambda = -3/l^2 $. The Lagrangian density $\mathcal{L(F)}$ is a function of $\mathcal{F}=\frac{1}{4}\mathcal{F}^\textit{ab}\mathcal{F}_\textit{ab}$ with $\mathcal{F}_\textit{ab}=\partial_\textit{a}\mathcal{ A}_\textit{b} - \partial_\textit{b}\mathcal{ A}_\textit{a}$ and the solution we are interested, has the form
 
\begin{equation} \label{L_term}
\mathcal{L}(F) = F \exp[{{-k/e}~(2e^2F)^{\frac{1}{4}}}],\,\,\,\,\text{with}~~~~F=\frac{e^2}{2r^4}.
\end{equation}
where $e$ is NED charge and $k>0$ is constant.
	 %$ \mathrm{F}_{ab} = \partial_a A^b -\partial_b A^a $ is the electromagnetic field tensor and $ A_b=-\frac{\mathrm{Q}}{\textit{rdt}}$ is the vector potential.
To obtain our black hole solution, we assume the static spherically symmetric metric of the form \cite{Kumar:2020cve} 
\begin{equation}\label{Fr}
ds^2 = -{f}(r)dt^2+\frac{1}{f(r)}dr^2+r^2 d\Omega,
\end{equation}
where the function $\textit{f}(r)$ is the metric function to be determined and $\textit{d}\Omega = \textit{d}\theta^2+\sin^2\theta\textit{d}\phi^2$.
The Nambu-Goto action of strings evolving in spacetime is given by \cite{Letelier:1979ej,Singh:2020nwo}
  \begin{equation}\label{act}
 {S_M}= \int_ \Sigma \mathcal{M}\,(\gamma)^{-\frac{1}{2}} \,d\lambda^0 d\lambda^1 = \int _ \Sigma \mathcal{M} \left[-\frac{1}{2}\Sigma_{\mu\nu
 }\Sigma^{\mu\nu}\right]^\frac{1}{2} d\lambda^0 d\lambda^1,
 \end{equation}
 where the world sheet is parameterized by spacelike and timelike parameters \cite{synge} represented by $\lambda^0$ and $\lambda^1$ and $\mathcal{M}$ is a dimensionless positive constant which characterises each string. The quantity $\gamma$ is determinant of $\gamma _{ab}$ given by 
 \begin{equation}\label{gamma} \gamma_{ab}=g_{\mu\nu}\frac{\partial\textit{x}^\mu}{\partial\lambda^a}\frac{\partial\textit{x}^\nu}{\partial\lambda^b}.   
 \end{equation}
 The movement of a string in time sweeps out area in two dimension which is termed as it's world sheet $\Sigma$ \cite{nambu} and has associated with it a bivector given by \cite{Letelier:1979ej}
 \begin{equation}\label{sigma}
 \Sigma^{\mu\nu}= \epsilon^{ab}\frac{\partial\textit{x}^\mu}{\partial\lambda^a}\frac{\partial\textit{x}^\nu}{\partial\lambda^a},
 \end{equation}    
 where $\epsilon^{ab}$ is the Levi-Civita tensor in two dimensions, which is anti-symmetric in the indices $a$ and $b$ given by $\epsilon^{01}=-\,\epsilon^{10}=1$.
 By varying the action (\ref{action}) with respect to $g_{ab}$, we obtain the following equations of motion
 \begin{equation}\label{Gab}
\textit{G}_{ab}-\frac{3}{\textit{l}^2} g_{ab}=\mathcal{T}_{\textit{ab}}=2\left(\frac{\partial\mathcal{L(F)}}{\partial\mathcal{F}}\mathcal{F}_{\textit{ac}}\mathcal{F}^{\textit{c}}_{\textit{b}}-g_{ab}\mathcal{L(F) }\right) + T_{ab}^\mathrm{cs},  
\end{equation}
and \begin{equation}\label{Lf}
\Delta_\textit{a}\left(\frac{\partial\mathcal{L(F)}}{\partial\mathcal{F}}\mathcal{F}_{\textit{ac}}\right) = 0,
\end{equation}
where $\textit{G}_{ab}$ is the Einstein tensor and 
\begin{equation} \label{tab}
T_{ab}^\mathrm{cs}=2\partial L/\partial g_{ab}=\rho\,(\gamma)^{-1/2}\Sigma^h_a\Sigma_{hb},    
\end{equation}
 is energy-momentum tensor of clouds of strings \cite{Ghosh:2014dqa}. The associated quantity $\rho$ is the proper density of the string cloud \cite{Ghosh:2014dqa} and $\rho \, (\gamma^{1/2}) $ is the gauge invariant density. The conservation of energy and momentum yields the equation
\begin{equation}
\nabla_a\,T_{ab}\,=\,0,
 \end{equation}
which leads to
\begin{equation}
\partial_a\,(\sqrt{-g}\,\rho\,\Sigma^{ab})\,=\,0.
\end{equation}
\begin{figure}
	\begin{tabular}{c }
		\includegraphics[width=0.5 \textwidth]{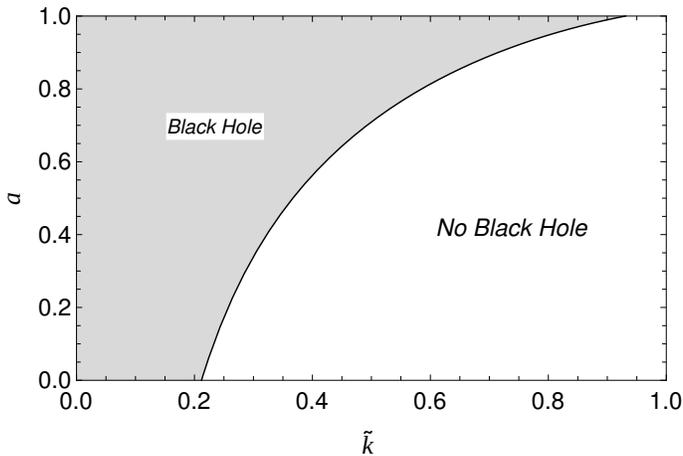}
	\end{tabular}
	\caption{The parameter space ($a$, $\tilde{k}$) for the existence of black holes. The black solid line corresponds to the extremal black holes.} 
	% one horizon where $ {a_0}=0.5594$ (left) and   %${k_0}=0.5039$ (right) of the Black Hole.}
	\label{fig:Bpotential1}
\end{figure}
For a spherically symmetric solution the density $\rho$ and bivector $\Sigma _{ab}$ are functions of radial and temporal components, hence the energy momentum tensor for a string cloud is \cite{Letelier:1979ej}
\begin{equation}\label{tt}
T^{t}_{t}\,=\,T^{r}_{r}\,=\,\frac{a}{r^{2}}.  
\end{equation}
On using Eqs. (\ref{Fr}), (\ref{Gab}) and (\ref{tt}), the $(r,r)$ component of Einstein equations can be integrated to  
\begin{equation}\label{metric}
f(r) = 1-\frac{2Me^{-k/r}}{r}+\frac{r^2}{l^2}-a,
\end{equation} 
where $M$ is the integration constant related to the mass of the black hole and $k$ is related to charge $e$ via $e^2=2Mk$.
Clearly solution (\ref{metric}) in the absence of clouds of strings background $(a\,=\,0)$, exactly goes over to solution in \cite{Kumar:2020cve} and encompasses the Schwarzschild black holes \cite{Ghosh:2014dqa, Bekenstein:1973mi,Schwarzschild:1916uq} when ($1/l^2\,=\,k\,=\,a\,=\,0$). 

We know that the metric (\ref{Fr}) with (\ref{metric}), in the absence of clouds of strings, is a regular or nonsingular black hole \cite {Kumar:2020cve}. However, with the background strings of clouds, we find the invariant Ricci scalar  $R=R_{ab}\,R^{ab}$, ($R_{ab}$ Ricci tensor) and the $K=R_{abcd}\,R^{abcd}$($R_{abcd}$ Reimann tensor) takes the form
\begin{equation}
    R~=\frac{2a}{r^2}+~\frac{2~M~ k^{2}~e^{-k/r}}{r^5},  
\end{equation}
\begin{equation*}
  K=\frac{4a^2}{r^4}+\frac{16e^{-k/r}Ma}{r^5}+      
    \end{equation*}
\begin{equation}
    \frac{4M^2 e^{-k/r} (k^4-8k^3r+24k^2r^2-24kr^3+12r^4)}{r^{10}}
    \end{equation}
which obviously diverges, when $a, M\neq0$ and $r\to0$. Thus, the spacetime  with the background strings of clouds is no more  regular, but it becomes singular.
\paragraph{Energy Conditions}
Next, we check the status of the various energy conditions using the prescription of Hawking Ellis \cite{Hawking:1973uf,Ghosh:2008zza,Kothawala:2004fy}. The Einstein equations governing the stress energy tensor $T_{\mu\nu}$ is given by the Eq.~(\ref{Gab}), which leads to
\begin{equation}\label{pr}
\begin{aligned}
\rho= \frac{2\,M\,k\,e^{-k/r}}{r^4} +\frac{a}{r^2} = -P_{r},\\
P_{\theta} = P_{\phi}= -\frac{M~k~e^{-k/r}(k-2r)}{r^5}. 
\end{aligned}
\end{equation}

 The weak energy condition (WEC) demands that $T_{\mu\nu}t^{\mu}t^{\nu}\geq 0$ everywhere, for any timelike vector $t^{\mu}$, which is equivalent to
 
 \begin{equation}
 \rho \geq 0, \;\; \rho + P_{i} \geq 0 \, (i=r,\; \theta,\; \phi),\\
 \end{equation}
and hence    
\begin{equation}\label{r}
    \rho+P_{\theta}=\frac{a}{r^2}-\frac{Mk(k-4r)e^{-k/r}}{r^5}.
\end{equation}
Hence, the weak energy condition is satisfied when $ k\leq4r$.
Next, the null energy condition (NEC) requires that $T_{\mu\nu}t^{\mu}t^{\nu}\geq 0$ in the entire spacetime, for any null vector $t^{\mu}$. The null energy condition demands $\rho+P_{r}\geq0$ and $\rho+P_{\theta}\geq0$. The former becomes identically zero and $\rho+P_{\theta}$ becomes positive for $ k\leq4r $. Hence the null energy condition is also satisfied for $k\leq 4r$.

Finally, the strong energy condition (SEC) states that $T_{\mu\nu}t^{\mu}t^{\nu}\geq 1/2T^{\mu}_{\mu}t^{\nu}t_{\nu}$ globally, for any timelike vector $t^{\mu}$, which requires
\begin{equation}
    \rho+P_{1}+2~P_{2}\geq 0.
\end{equation}
We find that 
\begin{equation}
    \begin{aligned}\label{st}
    \rho+P_{1}+2~P_{2} = \frac{a}{r^2}-\frac{3Mke^{-k/r}\left(k-\frac{8r}{3}\right)}{r^5}.
\end{aligned}
\end{equation} 
     Thus, the strong energy condition is satisfied for $k\leq3/8~r$. We find that the energy conditions are satisfied when $r>k/4$  and hence they are violated in the deep core.

\paragraph{Extremal black holes}
In order to discuss the properties of the solution
(\ref{metric}) we rewrite it as
\begin{equation}\label{metr}
f(r) = 1-\frac{2me^{-\tilde{k}/x}}{x}+x^2-a,
\end{equation} 
such that $ x=r/l$, $m=M/l$ and $\tilde{k}=k/l$. Thus the black hole (\ref{metric}) is characterised by three parameters $ M$, $k$ and $a$. The metric (\ref{Fr}) is singular at the points where $ f(x)\,=\,0$ and it corresponds to coordinate singularity, signifying the presence of event horizon. The numerical solution of $ f(x)=0$ reveals existence of maximum two roots as shown in Fig.~\ref{fig : MvsX_m}. It is possible to find critical value $m_0$, for a given $a$ and $\tilde{k}$, such that for $m > m_0$ there exists two roots of $ f(x)\,=\,0$ viz. ${\pm}\,x\,$ corresponding to Cauchy horizon ($x_-$) and event horizon ($x_+$) (cf. Fig.~\ref{fig : MvsX_m}). When $m=m_o$ the two horizon degenerate on ($x_+=x_-=x_0 $) and we have extremal black holes. Similarly, for given $\tilde{k}$ and $m$, we have found critical value of $a$ such that $a=a_0$ corresponds to extremal black hole (cf. Fig.~\ref{fig:Bpotential}) and $a>a_0$ corresponds to regular black holes with two horizons ($x_{\pm}$). In Fig.~\ref{fig:Bpotential1}, we have depicted the parametric space ($a$, $\tilde{k}$) for $m=0.3$, in which the solid black curve is corresponding to the extremal black holes. The grey region is showing black holes with two horizons whereas the white region corresponds to no black hole solutions. In absence of clouds of strings parameter ($a=0$), we only obtain critical values $m_0$ and $\tilde{k}_0$ which correspond to extremal black hole at $m=m_0$ ($\tilde{k}=\tilde{k}_0$) and regular black hole with two horizons for $m>m_0$ ($\tilde{k}<\tilde{k}_0$) \cite{Kumar:2020cve}. Interestingly the radius of an extremal black hole is slightly smaller when we introduce the clouds of strings parameter. 
 
\begin{widetext}

\begin{figure}[h!h!] 
\begin{tabular}{c c }
\includegraphics[width=0.5\textwidth]{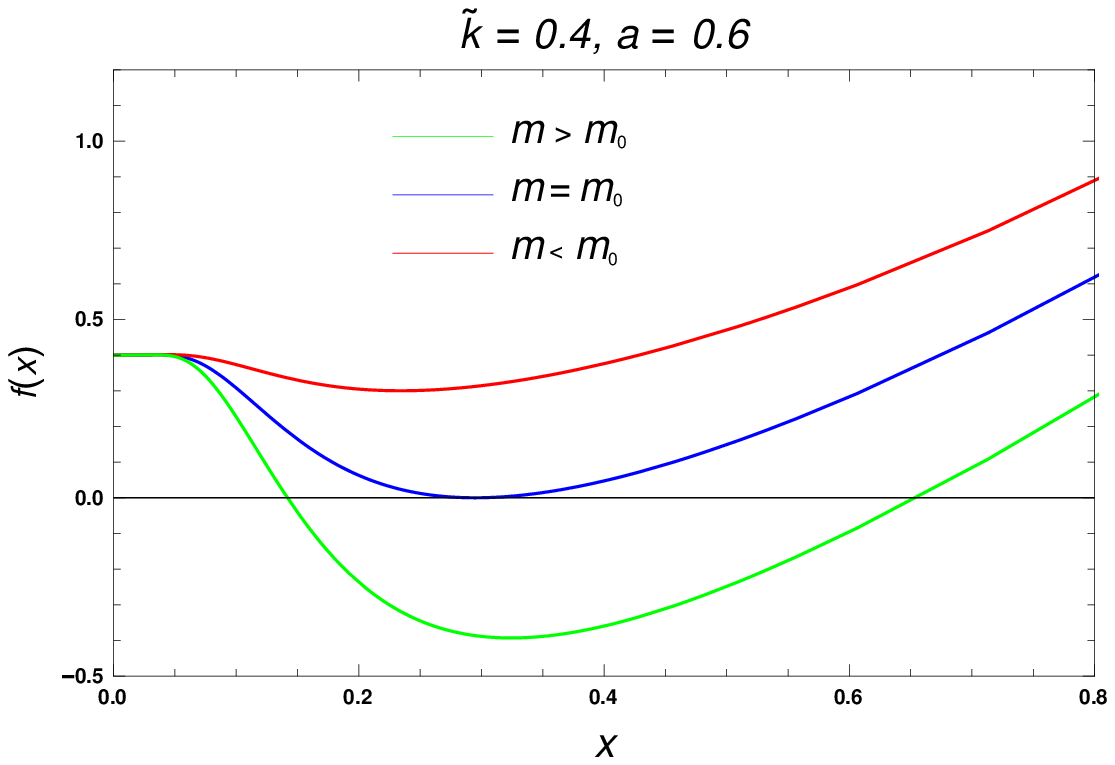}
\includegraphics[width=0.5\textwidth]{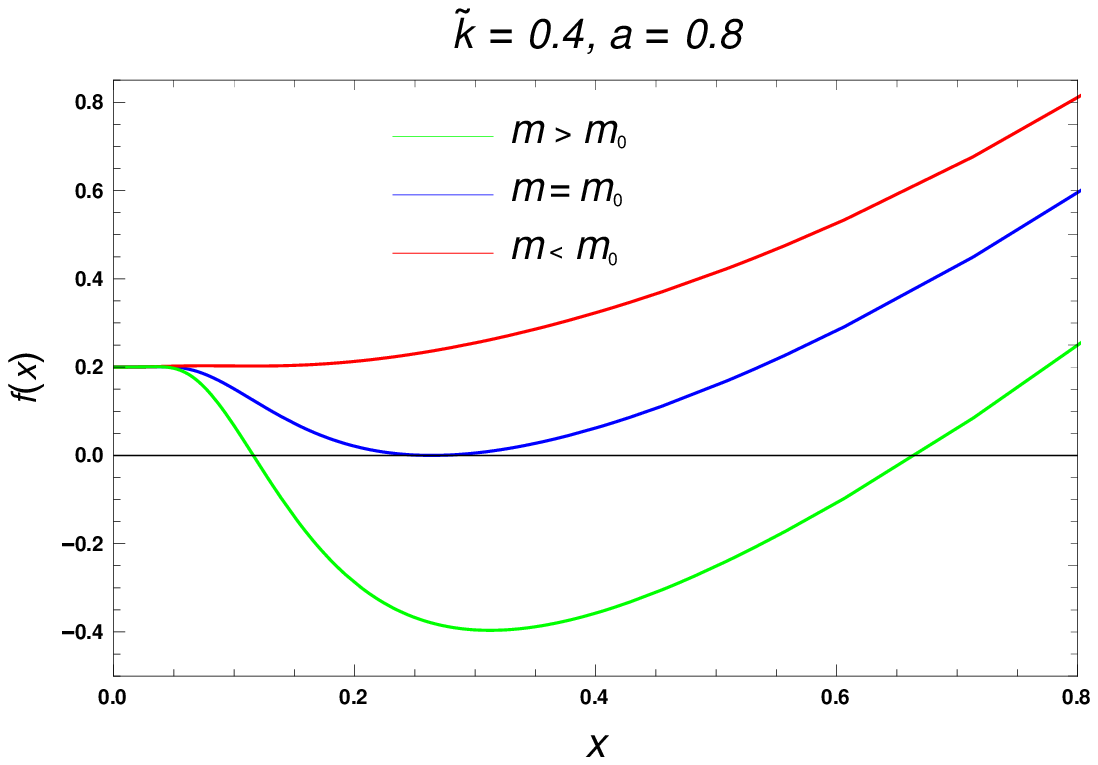}
\end{tabular}
\caption{ The metric function $f(x)$ vs $x$. When $m= m_0=0.5039$ (left) and   $m=m_0=0.16139$ (right) corresponding to extremal black hole with degenerate horizon. The black hole with Cauchy and event horizons exist when $ m > m_0 $.}
\label{fig : MvsX_m}
\end{figure}

\begin{figure}
\begin{tabular}{c c c c}
\includegraphics[width=0.5 \textwidth]{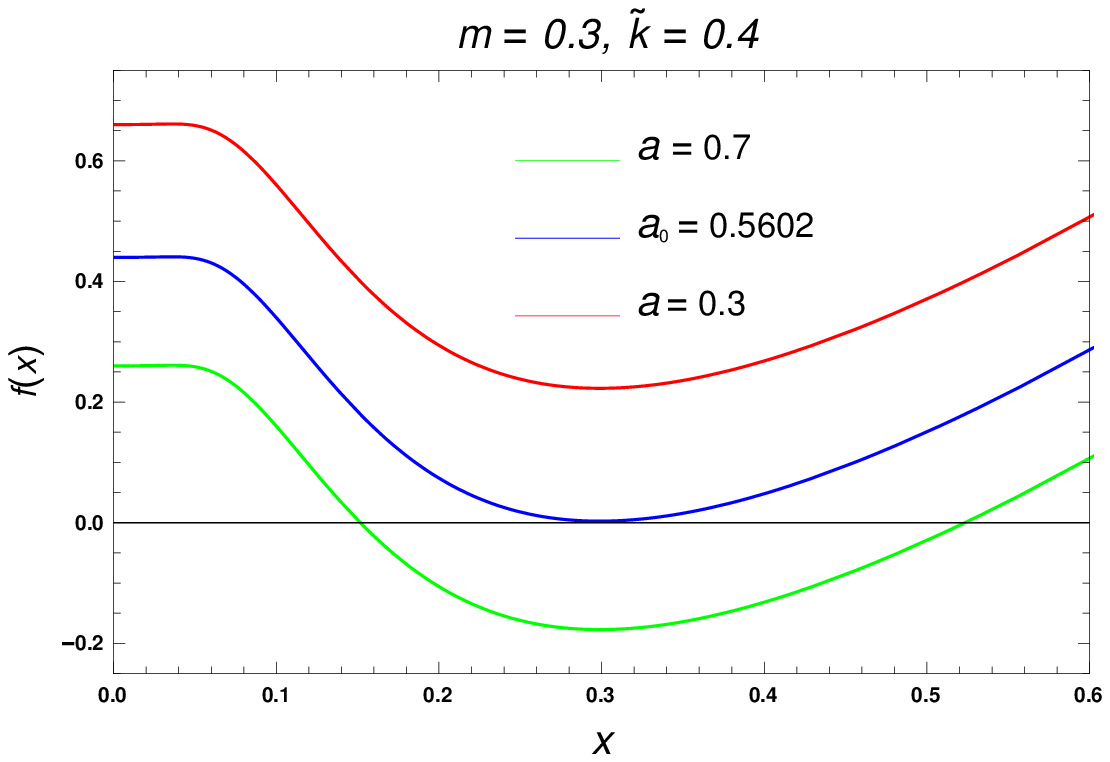}
\includegraphics[width=0.5 \textwidth]{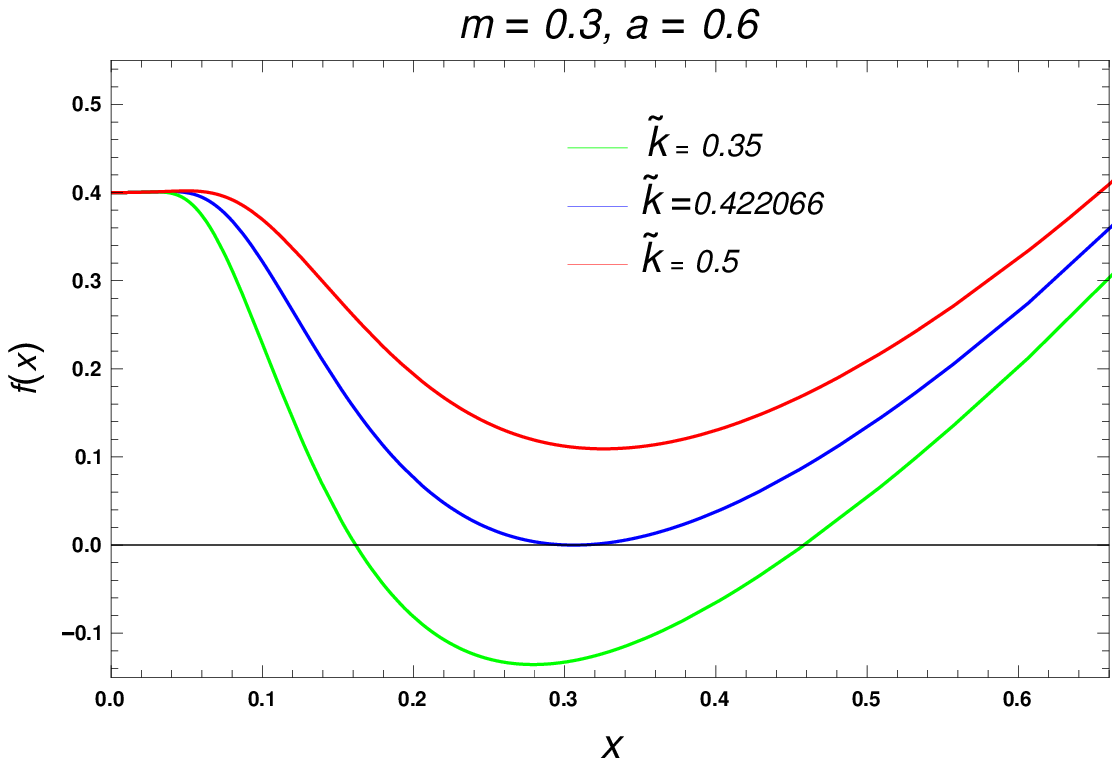}
\end{tabular}
\caption{The metric function $f(x)$ vs $x$ showing extremal black hole with degenerate horizon for $ a=a_0=0.5602$ (left) and   $\tilde{k}=\tilde{k}_0=0.42206$ (right) and the black hole with Cauchy and event horizons exist for $ a > a_0 $ and $ \tilde{k} <\tilde{ k}_0 $.} 
	% one horizon where $ {a_0}=0.5594$ (left) and   %${k_0}=0.5039$ (right) of the Black Hole.}
\label{fig:Bpotential}
\end{figure}

\end{widetext}

 The existence of extremal black hole confirms that the black hole undergoing evaporation will result in stable black hole remnant characterised by mass ($m_0$) and radius size ($x_0$), can be determined by solving systems \cite{Kumar:2020cve,Tzikas:2018cvs}
\begin{equation} \label{fx}
f(x)\,=\,0\,=\,\frac{\partial f(x)}{\partial x}\vert_{x=x_o}.
\end{equation}
Solving Eq. (\ref{fx}), we obtain
\begin{equation}\label{x0}
x_o=\frac{1}{9\upsilon}\left[\upsilon^2 + \tilde{k}\upsilon + \tilde{k}^2-9(1-a)\right],
\end{equation}
and 
\begin{equation}\label{m0}
m_o=e^{ 9\tilde{k}\upsilon/\mu} \left[\frac{81(1-a)\upsilon+\left(\upsilon^3+\tilde{k}\upsilon+\tilde{k}^2-9(1-a)\right) \mu }{81\upsilon}\right],   
\end{equation} where 
$\upsilon=\left[\tilde{k}^3-108(1-a)\tilde{k}+9\sqrt{3}\sqrt{\left(1-a\right)\beta}\right]^\frac{1}{3}$,\\ $\mu=\upsilon^2+\tilde{k}\upsilon+\tilde{k}^2-9(1-a)$ and $\beta = \tilde{k}^4-3\left(1-a\right)^2+47\left(1-a\right)\tilde{k}^2$. 
The above results, in the limit $a \to 0 $ reduce to those obtained in Ref. \cite{Kumar:2020cve}.
%\vspace{5cm}
\section{Black hole thermodynamics}\label{Sec3}
\begin{widetext}
	
\begin{figure}
	%	\centering
	\begin{tabular}{c c}
	\includegraphics[width=0.5\textwidth]{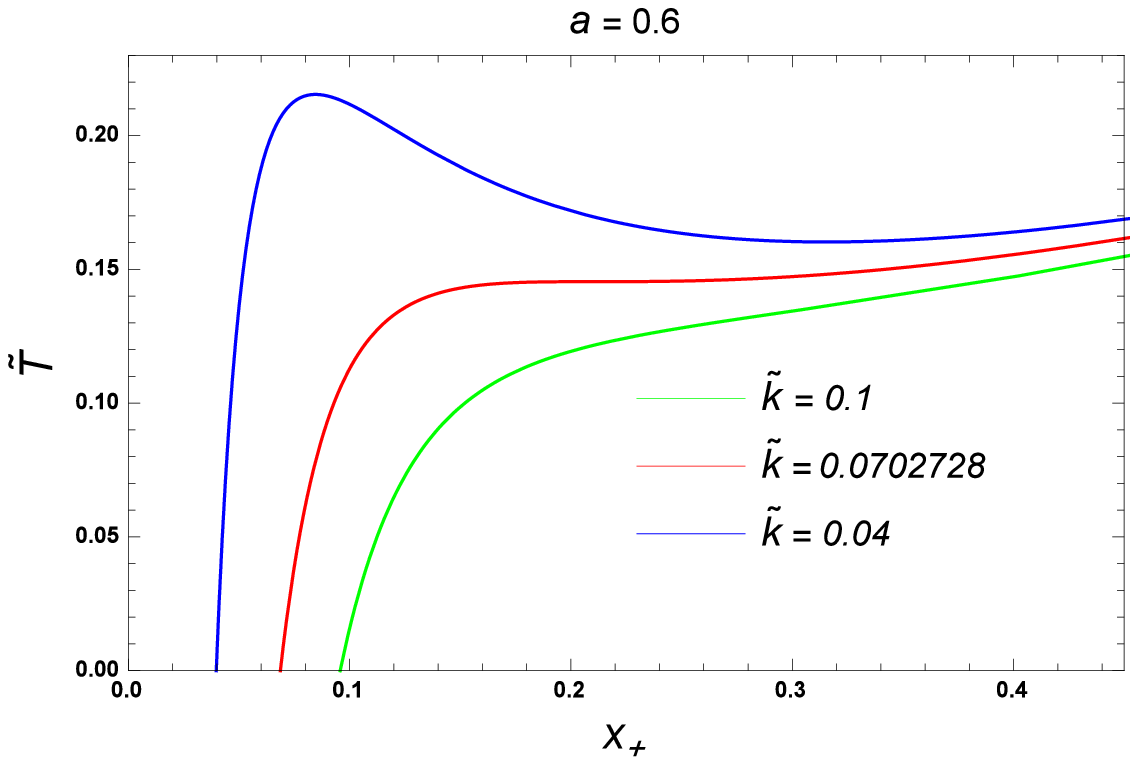}
	\includegraphics[width=0.5\textwidth]{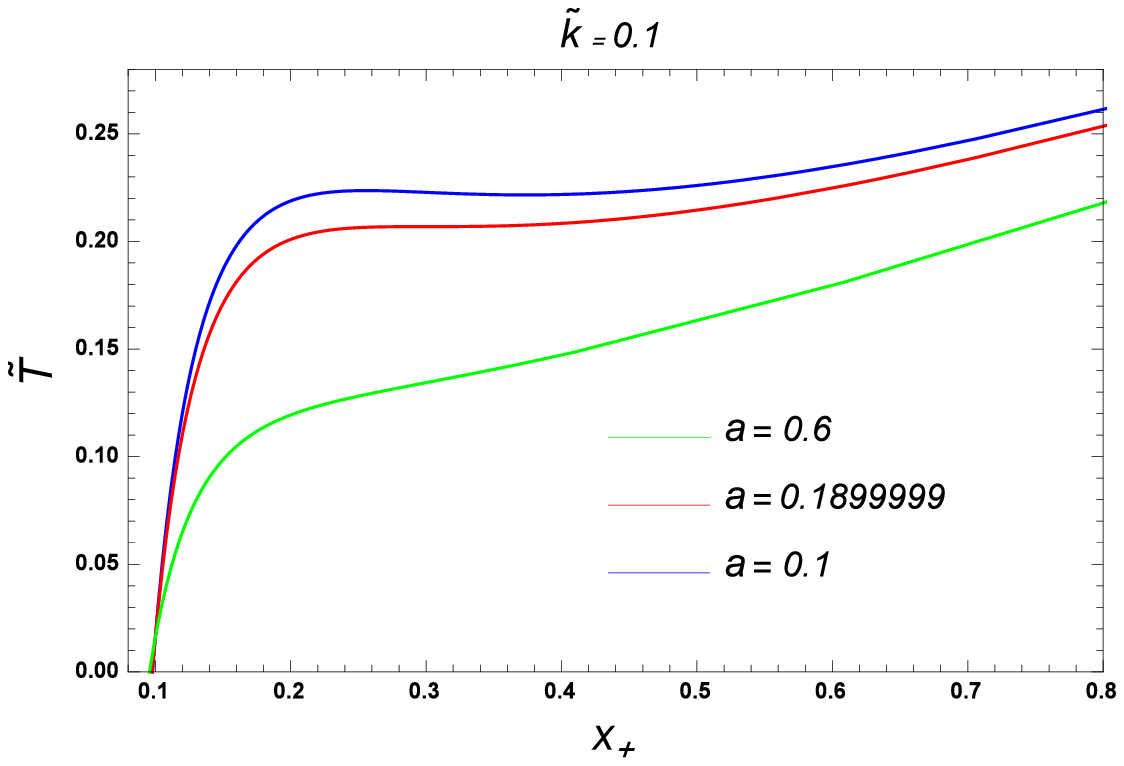}
    \end{tabular}
	\caption{The Hawking temperature $\tilde{T}$ vs horizon $x_+$ for $a=0.6$  varying $\tilde{k}$ with $\tilde{T}_{max}=0.215369 $ , $\tilde{T}_{min}=0.160236 $ for $\tilde{k}=0.04$  (left),  and for $\tilde{k}=0.1$ varying $a$ with $\tilde{T}_{max}=0.223639 $, $\tilde{T}_{min}=0.221622 $ for $a=0.1$  (right). For $a=0.6$,  $\tilde{T}_{max}$ = $\tilde{T}_{min}$ at $x=0.210$ for $\tilde{k}=0.0702728$ whereas for $\tilde{k}=0.1$, $\tilde{T}_{max}$ = $\tilde{T}_{min}$ at $x=0.3$ for $a_c=0.1899999$.   } 
	\label{Temp}
	\end{figure}

\end{widetext}

Black hole thermodynamics conceived as a discipline, was formulated with the advent of quantum field theory in curved spacetime which suggested analogies of thermodynamic quantities like internal energy, temperature and entropy with black hole mass, surface gravity and event horizon area, respectively \cite{Bekenstein:1973mi,Bardeen:1973gs,Bekenstein:1972tm,Bekenstein:1974ax}. The nomenclature in black hole chemistry, has been given in a new scientific era and is simply regarded as the thermodynamics of black holes in extended AdS spacetime, \cite{Kubiznak:2014zwa,Kubiznak:2016qmn} by demonstrating cosmological constant $\Lambda$ as dynamical thermodynamical pressure \cite{Teitelboim:1985dp,Brown:1987dd} via $P=3/8 \pi l^2$. This inclusion of $\Lambda$ as thermodynamical pressure ($P$) leads to interpretation of black mass as enthalpy $(H\,=\,m)$ instead of internal energy \cite{Dolan:2011xt,Dolan:2010ha}. Then the generalised first law of black hole thermodynamics reads as \cite{Kastor:2011qp,Kubiznak:2016qmn,Kubiznak:2014zwa,dr2,Kubiznak:2012wp}
 \begin{equation}\label{firstlw}
dm= TdS + VdP +\phi dk,
\end{equation} where $m$ $(S,k,P)$ is black hole mass function with dynamical variables namely the entropy $S$, the pressure $P$ and the magnetic charge parameter $k$. While the temperature $T$, the thermodynamical volume $V$ and the chemical potential $\phi$ are the conjugate variables of $S$, $P$ and $k$, respectively, which can be defined as \cite{Kastor:2011qp,Kubiznak:2016qmn,Kubiznak:2014zwa,dr2,Kubiznak:2012wp}
\begin{equation}
T=\left(\frac{\partial m}{\partial S}\right)_{k,P}, ~~~~\phi=\left(\frac{\partial m}{\partial k}\right)_{S,P}, ~~~~V=\left(\frac{\partial m}{\partial P}\right)_{k,S}.
\end{equation}
The cosmological term leads the Smarr relation \cite{Smarr:1972kt} to the following generalized form \cite{Zhang:2018hms,Kubiznak:2014zwa}
\begin{equation}\label{smarr2}
m=2TS-2PV+{\phi} k.
\end{equation}
The analysis of all thermodynamic quantities of the black hole is associated with the horizon $x_+$. We want to mention that $k$ and $e$ are not two independent parameters as in the case of Born-Infeld black holes \cite{Gunasekaran:2012dq}. Indeed, our solution is characterized by only one parameter $k$ apart from mass $M$, and in principle, we can work with only parameter $k$. 
The parameter $e$ is introduced if one wishes to compare our solution with the Reissner-Nordstrom metric via $e^2 = 2Mk$ for $r \gg k$.   Hence, there should only be the thermodynamic conjugate $\phi$ of parameter $k$ in the first law of black hole thermodynamics. The mass of the black hole is obtained by solving $f(x_+)=0$ as  
\begin{equation} \label{mass}
 m=H=\frac{x_+}{2}\,[e\,^{k/x_+}(1-a+x_{+}^2)].
\end{equation}
 The Hawking temperature of the black hole can be derived from the definition of surface gravity ($\kappa$) 
 \cite{Kubiznak:2016qmn} as follows
\begin{equation}\label{temp}
\tilde{T}=\frac{\kappa}{2\pi} =\frac{1}{4\pi x_+}[\,3x{_+}^2-\frac{\tilde{k}}{x_+}(1-a+x{_+}^2)+(1-a)].
\end{equation} 
${T} ={1}/{4\pi x_+}$ is the temperature of the Schwarzschild black holes.    
Obviously, the Hawking temperature of nonsingular-AdS black holes surrounded by clouds of strings is modified and it evident that the temperature is sensitive to the clouds of strings (cf. Fig.~\ref{Temp}).  It is evident from the  Fig.~\ref{Temp}, for a given $\tilde{k}$ ($a$), there is a phase transition with $0 < a < a_c$ (($0< \tilde{k}<\tilde{k}_c$)).  Fig.~\ref{Temp} shows that the Hawking temperature has local maxima and minima, respectively, at critical radii $x_{c1}=0.25578$ and $x_{c2}=0.31547$ when $a < a_c = 0.189999 $ (for $\tilde{k} = 0.1$) (cf. Fig.~\ref{Temp}). The critical value $a_c$ ($\tilde{k}_c$) for some given value of $\tilde{k}$ ($a$), can be obtained by solving $\partial \tilde{T}/ \partial x|_{x=x_c}=0$. The Hawking temperature plot is showing positive slope for small  black hole (SBH) and LBH whereas for the intermediate black hole (IBH) it has negative slope, mimicking the Van der Walls-like first-order phase transition between SBH--LBH \cite{Kubiznak:2012wp,Wei:2014hba}. When $a=a_c$ ($\tilde{k}=\tilde{k}_c$), the Hawking temperature's local maxima and minima merge and they increases monotonically when $a>a_c$ ($\tilde{k}>\tilde{k}_c$) (cf. Fig.~\ref{Temp}). Also, due to background clouds of string, the   $\tilde{T}_{max}$ increases. 

 The Bekenstein–Hawking entropy  of the black hole is obtained from the first law of black hole thermodynamics (\ref{firstlw}), which comes out to be
 	\begin{equation} \label{entr}
	\tilde{S}=\frac{A}{4}\left[{\frac{e~^{\tilde{k}/{x_+}}}{x_+}}\left(\tilde{k}+x_{+}\right)-\frac{\tilde{k}}{x{_+}^2} \mathrm{Ei} \left[\frac{\tilde{k}}{x_+}\right] \right],
    	\end{equation} 
where $A = 4\pi x_{+}^2$ and $\mathrm{Ei}$ is the exponential integral function.  Eq. (\ref{entr}) reduces to the usual area law ($S=A/4$) up to $\mathrm{Ei}$ corrections in the classical limit. It is interesting to note that the entropy (\ref{entr}) does not explicitly depend on the clouds of strings parameter $a$, but is endowed in the horizon $x_+$. 
The cosmological term permits the definition of the system's pressure and its conjugate quantity, the black hole volume, and we obtain pressure, volume and $\phi$ associated with the black holes as
\begin{equation}\label{PVphi}
P = \frac{3}{8\pi l^2},~~~
\tilde{V}=\frac{4}{3}e~^{\tilde{k}/{x_+}}\pi x_{+}^3,~~~ \bar{\phi}=\frac{e~^{\tilde{k}/{x_+}}}{2}\left[1-a+x_{+}^2\right].
 \end{equation}
In the classical limit ($a,k \to 0$), all above variables coincide with the conventional variables of a Schwarzschild AdS black hole \cite{Kubiznak:2012wp}, and in the case when only $a\,\to\,0$,  they go over to that of the nonsingular-AdS black holes \cite{Kumar:2020cve}.

\section{Stability and $P-V$ critically }\label{Sec4}
Here, we investigate the thermal properties and the phase structure of black holes in canonical ensemble (fixed NED charge). Hawking and Page \cite{Hawking:1982dh} investigated the thermodynamic properties of black holes in asymptotically AdS spacetimes. 
The thermodynamic stability of the black holes requires investigation of the Gibbs free energy  behaviour\cite{Kumar:2020cve,Tzikas:2018cvs}. We are interested in the regions where the free energy is negative, and identify where black holes are thermally favoured over the reference background. The free energy of the black hole is calculated from \cite{Kumar:2020cve,Tzikas:2018cvs}, 
\[G= m-\tilde{T}\tilde{S}\] which comes out to be complicated to present and hence depicted in Fig.~\ref{GvsX_a}

\begin{widetext}
	
	\begin{figure}
	\begin{tabular}{c c }
	\includegraphics[width=0.5\textwidth]{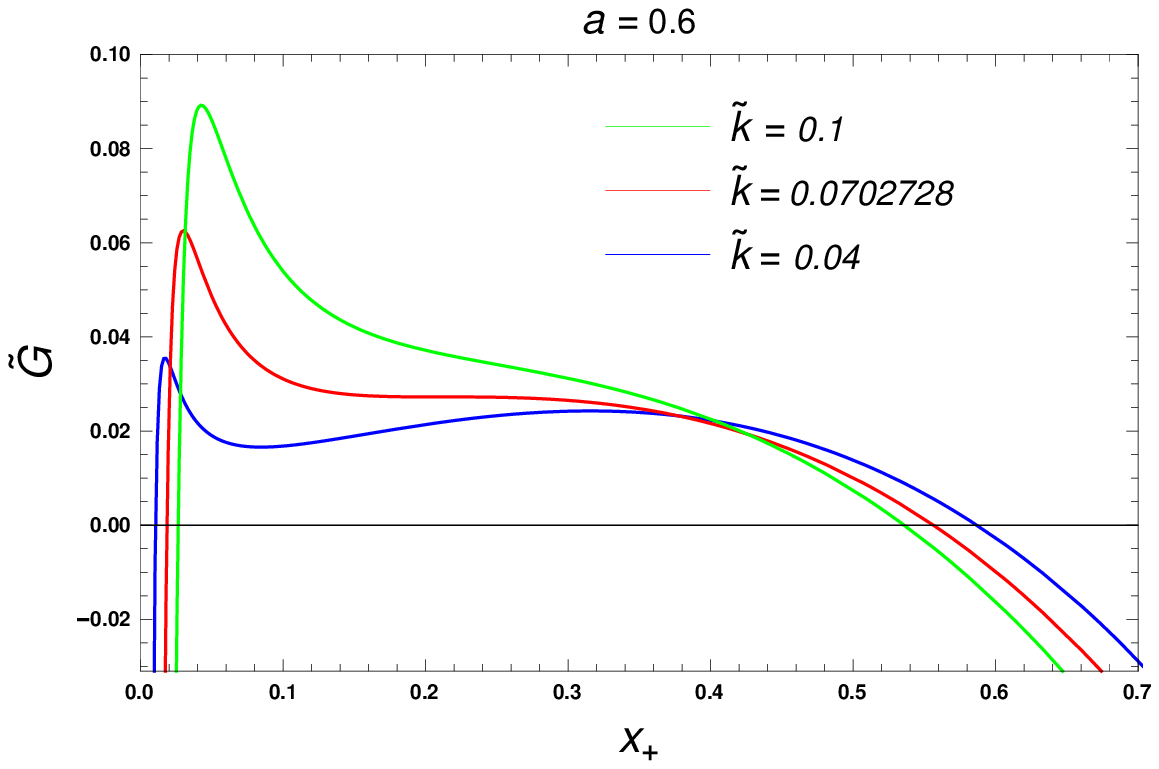}
	\includegraphics[width=0.5\textwidth]{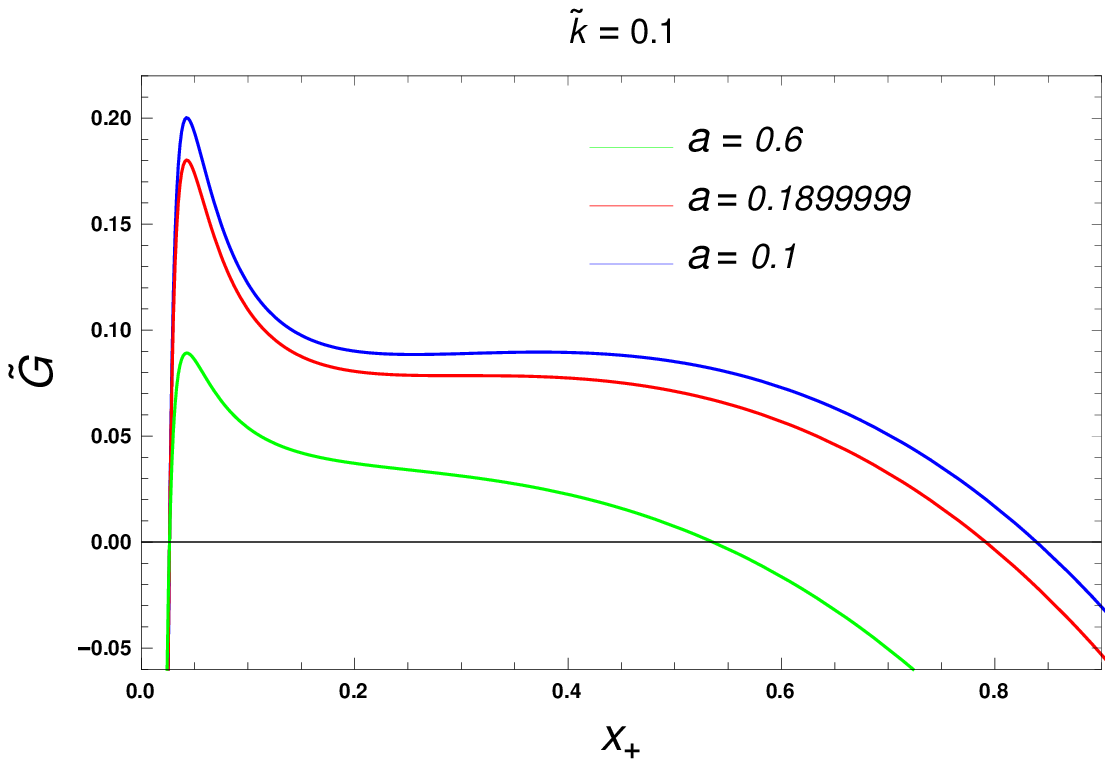}
	\end{tabular}
	\caption{The plot of Gibbs free energy $\tilde{G}$ vs horizon $x_+$ for nonsingular-Ads black hole surrounded by clouds of strings for different values of clouds of strings parameter $a$. }
	\label{GvsX_a}
\end{figure}
\begin{figure}
	\begin{tabular}{c c }
		\includegraphics[width=0.5\textwidth]{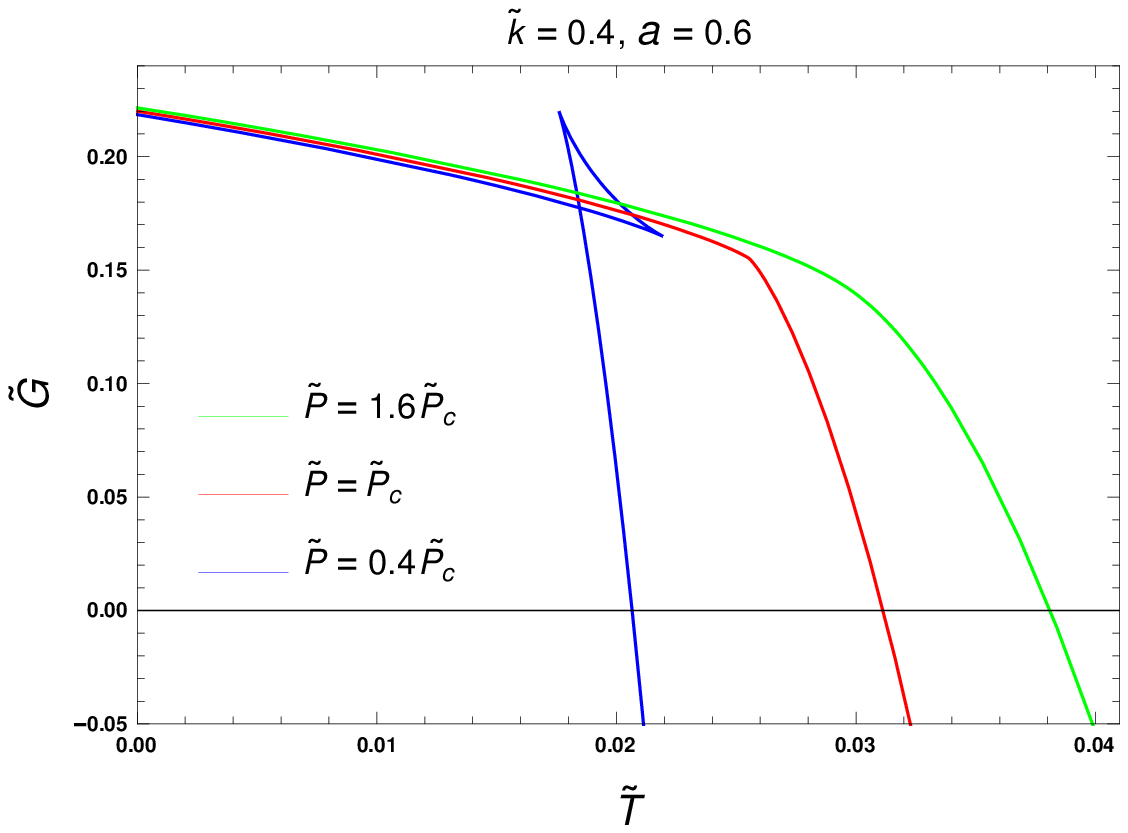}
		\includegraphics[width=0.5\textwidth]{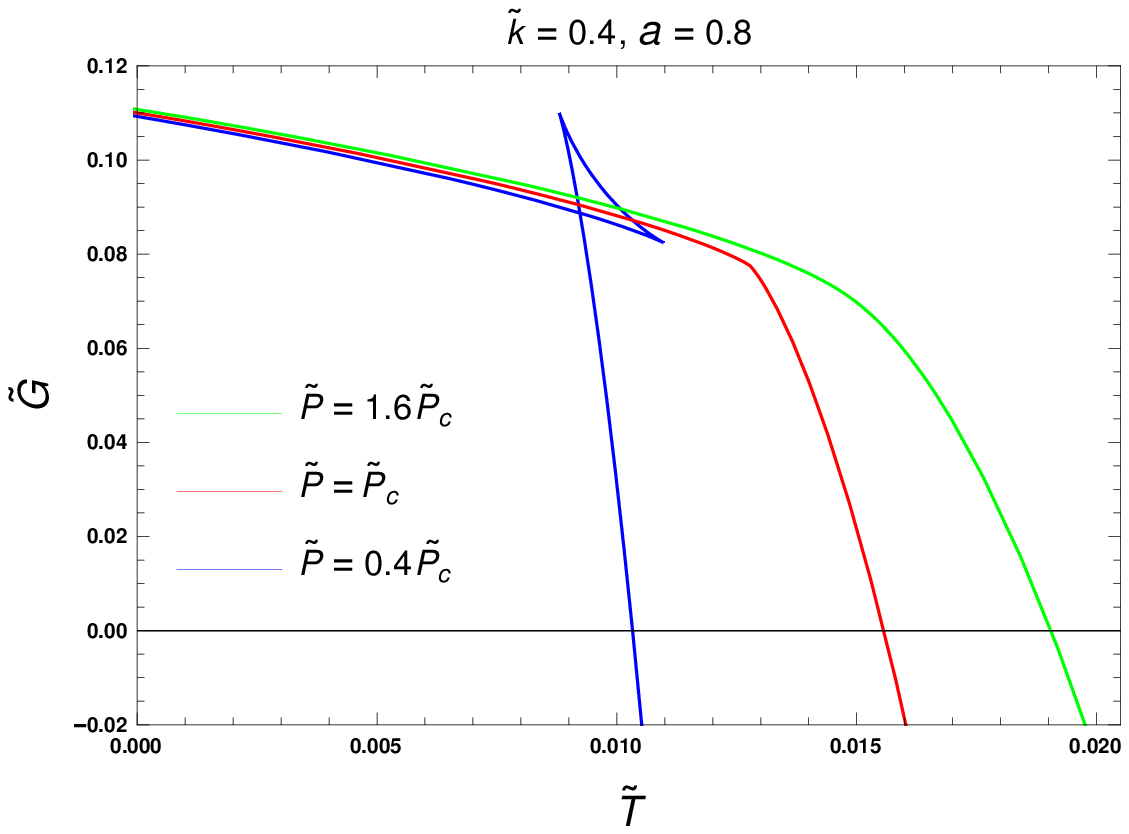}
	\end{tabular}
	\caption{The plot of Gibbs free energy $\tilde{G}$ vs temperature $\tilde{T}$ exhibiting a liquid-gas phase transition for $\tilde{P}<\tilde{P}_c$ and no transition for $\tilde{P}>\tilde{P}_c$ obtained for a particular value of $a$ ($\tilde{k}$) and corresponding critical pressure $\tilde{P}_c$. The swallowtail occurs at lower temperature for higher values of $a$.  	$\tilde{P}_c = 0.00368414$, $0.00184207$, respectively, for $a=0.6$, $0.8$. }
	\label{GvsT_a1}
\end{figure}

\end{widetext}

\begin{widetext}
	
 \begin{figure}
 \begin{tabular}{c c }
  \centering
 	\includegraphics[width=0.5\textwidth]{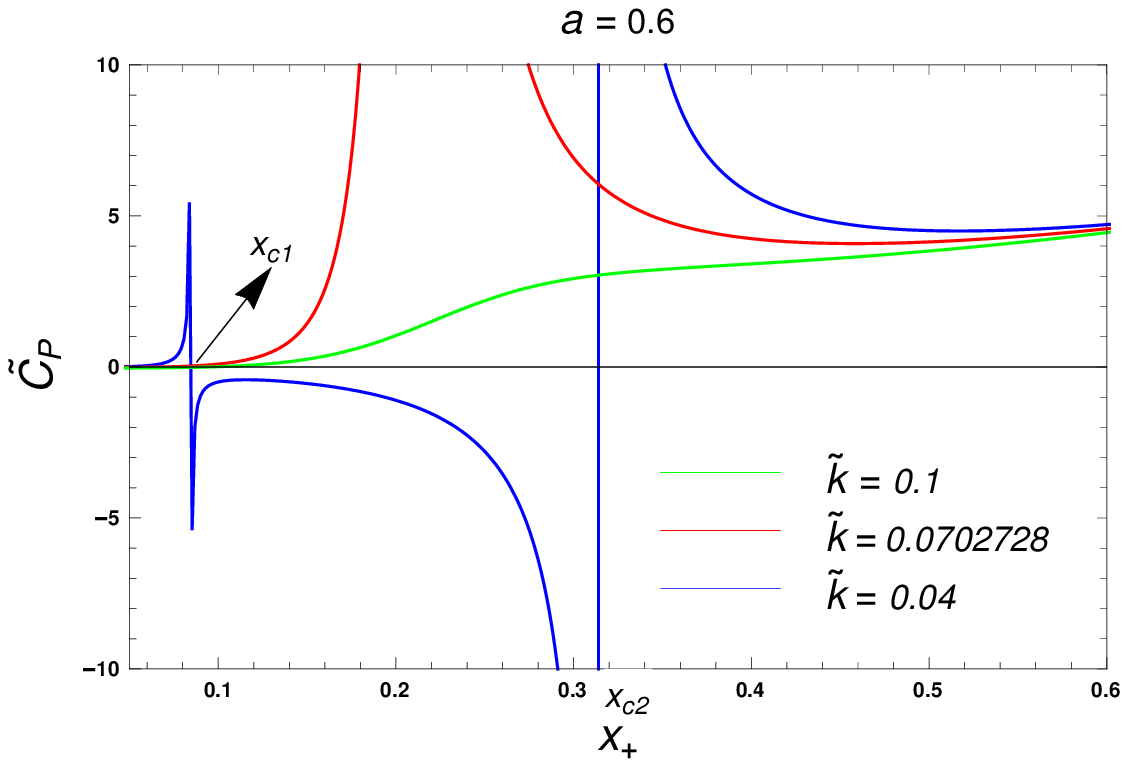}
  		\includegraphics[width=0.5\textwidth]{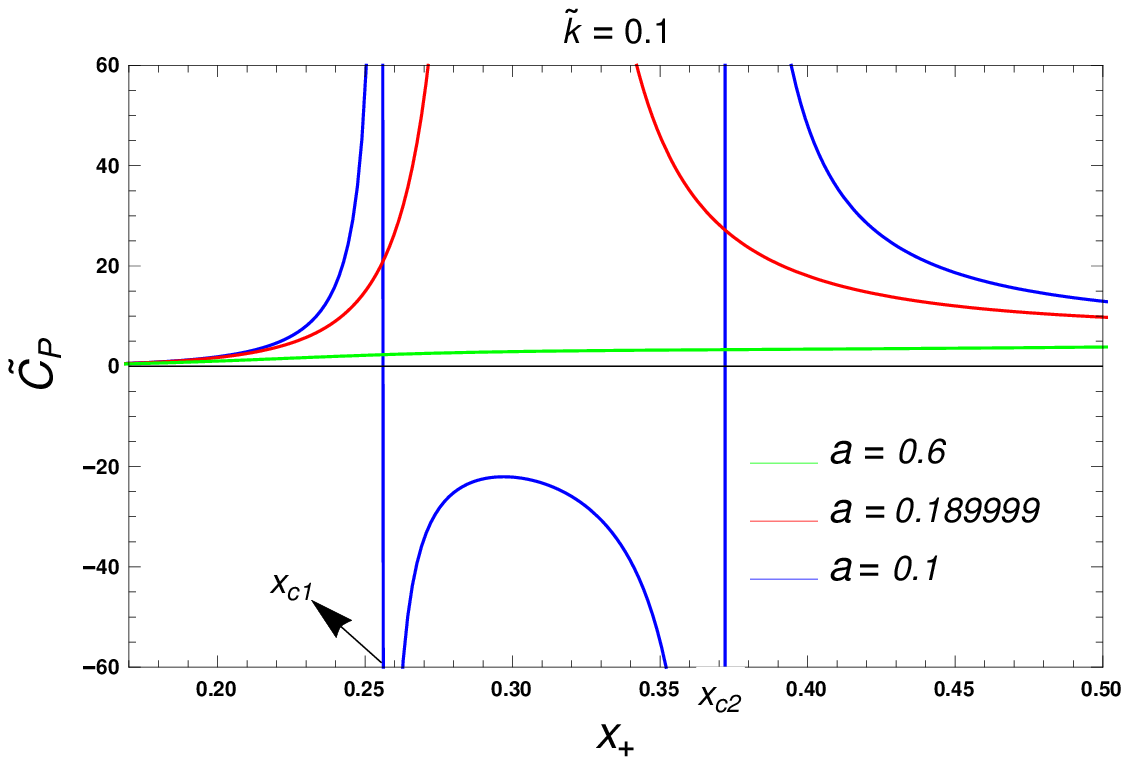}
 \end{tabular}
 	\caption{The plot of specific heat $\mathcal{\tilde{C_P}}$ vs horizon $x_+$ for different values of  parameter $a$ and NED parameter $\tilde{k}$.}
 	\label{Fig:CpvS}
 \end{figure}   
   
\end{widetext}

Since the global stability of the system is measured by the Gibbs free energy, its global minimum is estimated to be the preferred state of the black hole \cite{Kubiznak:2016qmn}.  Gibbs free energy plays a crucial role by identifying the first-order phase transition via the swallowtail behaviour in the  $\tilde{G}$-- $\tilde{T}$ plots (Fig.~\ref{GvsT_a1}), and $G$-$X_+$  plots (cf. Fig.~\ref{GvsX_a}) throw light on the SBH/LBH transition. As it is shown in  Fig.~\ref{GvsX_a}, the free energy $G_{+}$ for various $a$,  have local minimum and local maximum, respectively, at horizon radii $x_{c1}$ and $x_{c2}$ with $x_{c2}> x_{c1}$, which can be identified as the extremal points of the Hawking temperature shown in Fig.~\ref{Temp} and where the specific heat capacity $C_{+}$ diverges (cf. Fig.~\ref{Fig:CpvS}). The first-order phase transition occurs at $x_+=x_c$, where the free energy turns negative viz.,  $G_+<0$ for $x_{c}>x_{c2}$ . Thus the LBH, with horizon radii $x_+>x_c$, are thermodynamically globally stable. {\bf However, at minimal horizon radii, the Hawking temperature is negative and hence not physical for global stability.   The minimum value of the  $\tilde{G}$ decreases (increases), occur at the lower (higher) radius with increasing $a$ ($\tilde{k}$), and there will be no black hole for radii smaller than this. (cf. Fig.~\ref{GvsX_a})}.   

\begin{figure}
	\begin{tabular}{c c }
	\includegraphics[width=0.5\textwidth]{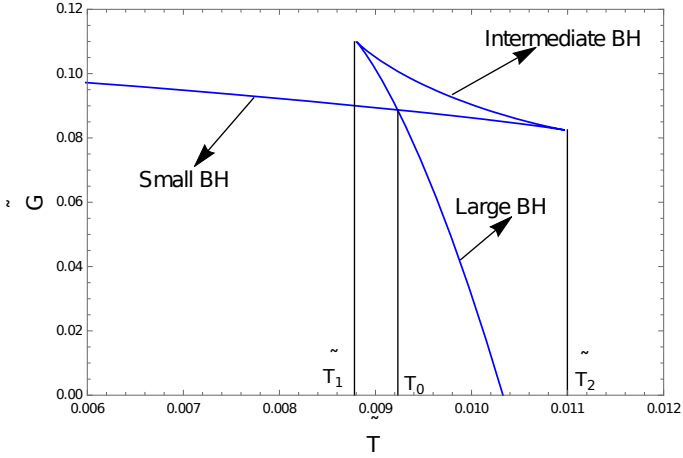}
\end{tabular}
	\caption{The generic plot of Gibbs free energy $\tilde{G}$ vs horizon $\tilde{T}$ for $\tilde{P}<\tilde{P}_c$.}
	\label{Figgt}
\end{figure}

 Fig.~\ref{GvsT_a1} shows the plot of the Gibbs free energy vs temperature. For the values of pressure below the critical pressure $\tilde{P}_c$, the $\tilde{G}-\tilde{T}$ diagram exhibits a swallowtail structure \cite{Nam:2019clw} where the Gibbs free energy of the black hole intersects with itself which is indicative of first-order phase transition between SBH and LBH (cf. Fig.~\ref{GvsT_a1}). By analyzing the Gibbs free energy, one can observe that at a pressure lower than the critical value ($\tilde{P}<\tilde{P}_c $), the Gibbs free energy exhibits a characteristic swallowtail behaviour which implies that SBH/LBH   first-order phase transition occurs. Indeed, for $\tilde{P}<\tilde{P}_c $, the $G$ -- $T$ graph exhibit three black holes states: SBH stable, intermediate BH (IBH) unstable and LBH stable (cf. Fig.~\ref{Figgt} ).  Further,
the SBH / IBH black hole transition occurs at the 
inflection point $T_2$ , the IBH/ LBH meet at $T_1$. 
Thus, LBH is the preferred state when $\tilde{T}\in[\tilde{T}_0,\tilde{T}_2]$, and for $\tilde{T}<\tilde{T}_0$, the preferred state is stable SBH. Thus, the SBH undergoes first-order phase transition to  LBH at $\tilde{T}=\tilde{T}_0$ (cf. Fig.~\ref{Figgt}). The second order phase transition is shown through the discontinuity of specific heat (c.f. Fig. 8) and through the cusp in $G-T$ plot when we take $P=P_C$.

After discussing the global thermodynamical stability of nonsingular-AdS black holes endowed with clouds of string, we next turn to the local stability or thermal stability.  The reason to consider local stability is that even when a black hole configuration is globally stable, it can be locally unstable \cite{Herscovich:2010vr}.  One can investigate the  thermodynamic stability from the specific heats $\mathcal{C_V}$ and $\mathcal{C_P} $ of the system   \cite{Kubiznak:2012wp}, where  $\mathcal{C_V}=T\left({\partial S}/{\partial T}\right)_{V.k}$ is the specific heat at constant volume and $\mathcal{C_P}=T\left({\partial S}/{\partial T}\right)_{P,k}$ is the specific heat at constant pressure \cite{Kubiznak:2016qmn}.
The black hole with positive specific heat is at least locally thermodynamical stable, whereas the negativity of specific heat signifies the thermodynamical instability of the black hole to the thermal fluctuations \cite{Sahabandu:2005ma,Cai:2003kt}. A phase transition is known to exist in the case when the specific heat diverges \cite{Tzikas:2018cvs}. The specific heat at constant volume, $\mathcal{C_V}$, is zero for our case. The specific heat at constant pressure, $\mathcal{\tilde{C_P}}$ is calculated to be 
\begin{equation}\mathcal{\tilde{C_P}}=2\pi e^{\tilde{k}/{x_+}}x_{+}^2\left[\frac{\left(1-a+3x_{+}^2\right)-\frac{\tilde{k}}{x_+}\left(1-a+x_{+}^2\right)}{\left(1-a\right)\frac{2\tilde{k}}{x_+}-\left(1-a-3x_{+}^2\right)}\right].
\end{equation}

The specific heat of the Schwarzchild AdS black holes can be obtained in the classical limits  $a=\; \tilde{k}\,=\,0$ \cite{Kubiznak:2012wp} whereas when only $a\,=\,0$ it resembles the specific heat of nonsingular-AdS black holes \cite{Kumar:2020cve}
\begin{equation}
\mathcal{\tilde{C_P}}=-2\pi x_{+}^2\left[\frac{1+3x_{+}^2}{1-3x_{+}^2}\right]
\end{equation}

To  analyse the behaviour, we depict the heat capacity  $\tilde{C_{P}}$ against $x_{+}$  in Fig.~\ref{Fig:CpvS} for different values of $\tilde{k}$ and $a$.  It turns out that there exists $a_c (\tilde{k}_c)$, for given $ \tilde{k} (a)$,  such that  the heat capacity is discontinuous at the critical radii $x_{c1}$ and $x_{c2}$,  corresponding to the local maxima and minima of Hawking temperature,  with  $x_{c1}<x_{c2}$, when $a<a_c$ ($\tilde{k}<\tilde{k}_c$) .  This signals a second-order phase transition \cite{Kumar:2020cve,Hawking:1982dh}.   As can be seen from the Fig.~\ref{Fig:CpvS}, we have three different cases: 

 \begin{itemize} 
         \item When $a<a_c$ ($\tilde{k}<\tilde{k}_c$), we have the second-order phase transition between SBH -- IBH and IBH--LBH such that the black holes with $x<x_{c1}$ (SBH) and $x>x_{c2}$ (LBH) are locally stable with positive specific heat whereas the black holes with $x_{c1}<x<x_{c2}$ having negative specific heat are locally unstable .   
     \item For $a>a_c$ ($\tilde{k}>\tilde{k}_c$), no phase transition occurs and the specific heat is always positive signifying the local stability of the black holes. 
     \item The $a=a_c$ case for $\tilde{k}=0.1$ corresponds to existence of two stable black holes (SBH and LBH) that coexist at the inflection point $x\approx0.2988$.    
 \end{itemize}

\begin{figure*}[t]
  \begin{tabular}{c c }
 		\centering
 		\includegraphics[width=0.5\textwidth]{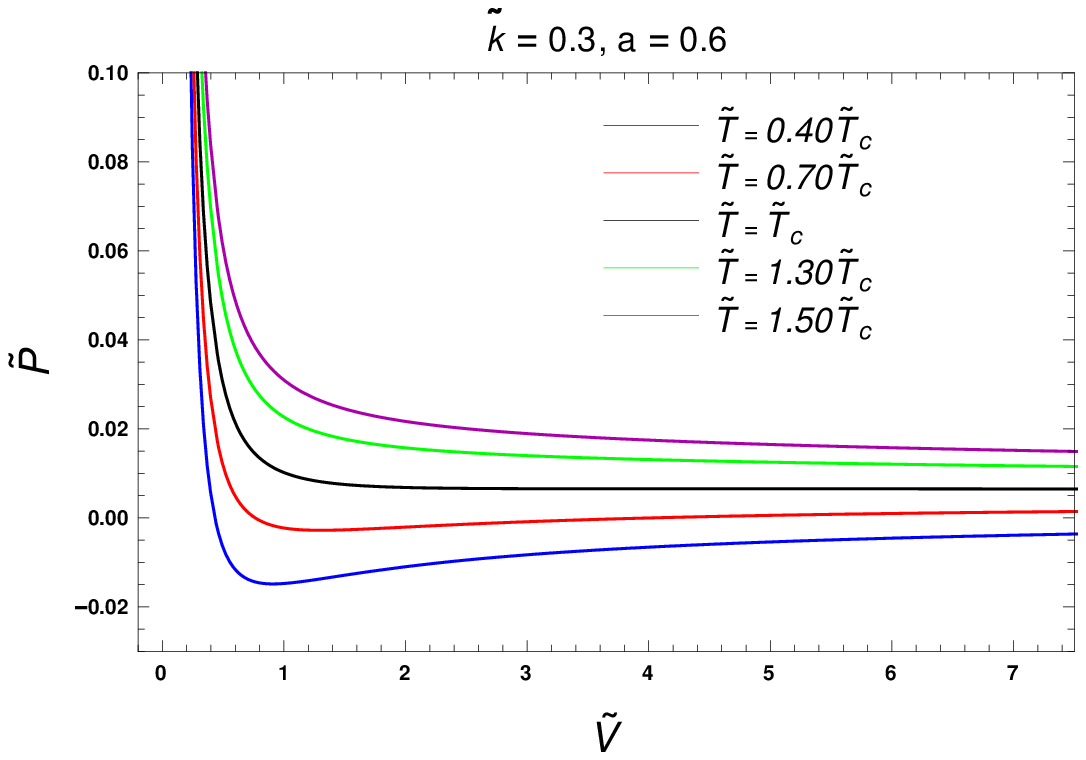}&
 		\includegraphics[width=0.5\textwidth]{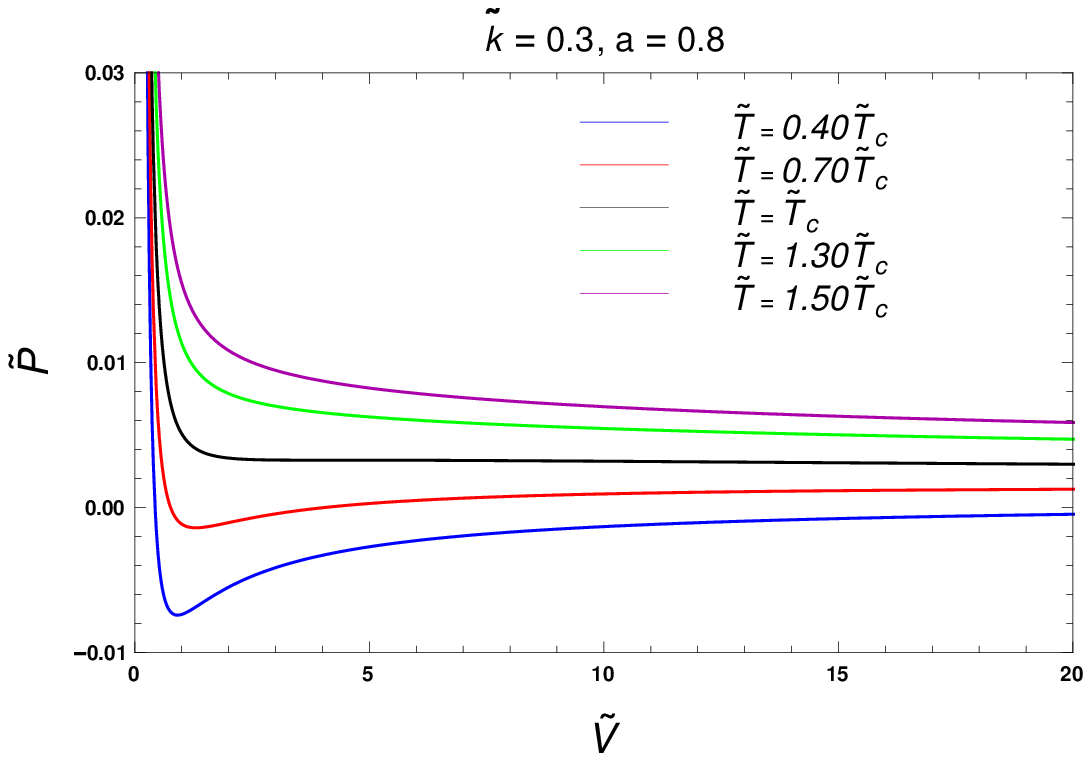}
 	\end{tabular}
 	\caption{The isotherms of black holes in $\tilde{P}$ - $\tilde{V}$ plane for  two different values of  parameter $a=0.6$ where $\tilde{T}_c=0.0340578$ (left) and $a=0.8$ where $\tilde{T}_c=0.0170289$ (right) with fixed $\tilde{k}=0.3$. The temperature of isotherms decreases from top to bottom}
 	\label{Fig:PvsV}
\end{figure*}

 The equation of state to investigate the $P-V$ criticality of the black hole, by using the equations of pressure, volume and temperature \cite{Kubiznak:2014zwa}, is found to be
\begin{equation} \label{state}
\tilde{\textit{P}} = \frac{3\tilde{\textit{T}}}{2\left[3x{_+}-\tilde{k}\right]} +\frac{3\left(1-a\right)\left[x_+-\tilde{k}\right]}{8\pi x{_+}^2\left[3x_+-\tilde{k}\right]},  
\end{equation}
where $\tilde{\textit{P}}=\textit{P}l^2$. The $\tilde{P}-\tilde{V}$ diagram depicted in Fig.~\ref{Fig:PvsV} plots the various isotherms of the equation of state Eq. (\ref{state}). For values of temperature $\tilde{T} $ higher than the critical temperature $\tilde{T}_c $, the plot shows the isotherms follow ideal gas behaviour. The isotherms undergo liquid-gas like phase transition \cite{Kubiznak:2014zwa} for $\tilde{T}\,<\,\tilde{T}_c$ governed by the Maxwell's equal area law \cite{Hyun:2019gfz} and have a point of inflection at $\tilde{T_c}$, which can be obtained from the equation
\begin{equation} \label{inflection}
\left.\frac{\partial \tilde{P}}{\partial x_{+}}\right|_{\tilde{T}}\,=\,0\,=\,\left.\frac{\partial^2 \tilde{P}}{\partial x_{+}^2}\right|_{\tilde{T}}. 
\end{equation}
The LBH are depicted by the branch with low pressure and SBH are shown by the branch with high pressure. The oscillating branch between SBH and LBH show the black hole is going through Van der Walls-like first order phase transition \cite{Hyun:2019gfz}. The Eq. (\ref{inflection}) yields the expression for the critical temperature $\tilde{T}_c$
\begin{equation}
    \tilde{T_c}\,=\dfrac{(1-a)(3x_{c}^2-5\tilde{k} x_{c}+\tilde{k^2})}{6\pi x_{c}^3},
\end{equation}
and further we can get the equation for critical pressure $\tilde{P}_c$ in terms of critical radius $x_c$ as 
\begin{equation}
    \tilde{P}_c\,=\,\frac{(1-a)(81\tilde{k}x_{c}-81\tilde{k}^2-26x_{c}^2)}{216\tilde{k}\pi (\tilde{k}-3x_{c})x_{c}^2}.
\end{equation}
Inserting the value of critical radius $x_c=3\tilde{k}$ in the above expressions for critical temperature, critical pressure and critical volume  we obtain the  following expressions 
\begin{equation}\label{PcVcTc}
    \tilde{P}_c=\frac{1-a}{216 \pi \tilde{k}^2},~~~~~
\tilde{V}_{c}=36\,e^{1/3}\pi \tilde{k}^3,~~~~~ \tilde{T}_{c}=\frac{13(1-a)}{162\pi\tilde{k}}.
\end{equation}
The universal constant  $\epsilon$ for the nonsingular AdS black hole endowed with clouds of string calculated as 
\begin{equation}\label{uc}
	\epsilon = \frac{\tilde{P}_c\,{\tilde{V}_c}^{1/3}}{{\tilde{T}_c}}=0.311, 
\end{equation}
which is slightly less than $3/8$ -- the value for  Van der Walls fluid \cite{Kumar:2020cve}.  It turns out that the temperature of isotherms decreases from top to bottom (cf. Fig.~\ref{Fig:PvsV}), and  ${\tilde{T}_c}$ decreases with increasing $a$.  Clearly, the quantities   $\tilde{P}_c$ $\tilde{V}_c$ ${\tilde{T}_c}$ are corrected because background clouds of strings.

\section{Critical Exponents}\label{Sec5}
 We analyse the thermodynamic quantities near the critical point with the help of critical exponents $\alpha$, $\beta$, $\gamma$, and $\delta$ \cite{Gunasekaran:2012dq}. These exponents do not depend on the physical system instead can depend on the dimension of the system or the range of the interaction. The exponents $\alpha$ and $\beta$ determine the behaviour of specific heat at constant volume \cite{Kubiznak:2012wp} and behaviour of the order parameter $\eta$ which is the difference between the volume of the gas phase and the volume of the liquid phase respectively. $\gamma$ explains the behaviour of isothermal compressibility $ \kappa_{\tilde{T}}$ and  $\delta$ governs the behaviour of the critical isotherm $ \tilde{P}\,-\,\tilde{P}_c $ \cite{Nam:2019clw}. 
 The critical exponents can be calculated by using following relations \cite{Kumar:2020cve,Tzikas:2018cvs}
 \begin{eqnarray}
\tilde{C}_{\tilde{V}} &=& \tilde{T} \frac{\partial \tilde{S}}{\partial \tilde{T}} \Big |_{\tilde{V}} \propto |t|^{-\alpha} \,,\;\;\;\;\; \label{crit_b}
\eta = \tilde{V}_l - \tilde{V}_s \propto |t|^{\beta} \,, \\ \label{crit_c}
\kappa_{\tilde{T}} &=& - \frac{1}{\tilde{V}} \frac{\partial \tilde{V}}{\partial \tilde{P}} \Big |_{\tilde{T}} \propto |t|^{-\gamma} \,,\; \label{crit_d}
|\tilde{P}-\tilde{P}_c|_{\tilde{T}=\tilde{T}_c}   \propto   |\tilde{V}-\tilde{V}_c|^{\delta},
\end{eqnarray}where $ \tilde{C}_{\tilde{V}} $ is the specific heat at constant volume, $\eta$ is the order parameter, $\kappa_{\tilde{T}}$ is the isothermal compressibility. $\tilde{V}_l$ and $\tilde{V}_s$ are the volumes of LBH and SBH. The specific heat at constant volume is independent of t hence the critical exponent $\alpha\,=\,0$ \cite{Hyun:2019gfz}. In order to study the critical exponents $\delta$, $\beta$ and $\gamma$ we define the following set of dimensionless quantities \cite{Kumar:2020cve,Nam:2019clw,Hyun:2019gfz}
\begin{equation}\label{dim}
 p = \frac{\tilde{P}}{\tilde{P_C}},~~~1+\epsilon = \frac{x_+}{x_c},~~~
 1+\omega = \frac{\tilde{V}}{\tilde{V}_c},~~~  1+t = \frac{\tilde{T}}{\tilde{T}_c},
\end{equation} 
where $\mid\epsilon\mid$, $\mid\omega\mid$, $\mid t \mid\, \ll 1$. 
Our analysis requires a relation between $\epsilon$ and $\omega$ which is obtained by using Eqs. (\ref{PVphi}) and (\ref{dim}) as
\begin{equation}\label{omega}
\omega=4\left(1-\frac{\tilde{k}}{2 x_c} \right)\epsilon.
\end{equation}  
 The equation of state given by Eq. (\ref{state}) in terms of dimensionless quantities defined by Eq. (\ref{dim}) becomes
\begin{equation}\label{state1}
p=1+\mathcal{ U}t-\mathcal{ V}t\epsilon-\mathcal{ W}\epsilon^3+\mathcal{ O}(t\epsilon^2, \epsilon^4),
\end{equation} 
where $~\mathcal{ U}=\dfrac{3}{ 2\left(3 x_c-\tilde{k}\right)}\frac{ \tilde{T}_c}{\tilde{P}_c}$,
~~$\mathcal{ V}=\dfrac{9 x_c}{2\, \left(3 \, \mathrm{x_c}-\tilde{k}\right)^2}\frac{ \tilde{T}_c}{\tilde{P}_c}$,
~and
\begin{widetext}
	
\begin{equation} \label{W}
\mathcal{ W}  =\dfrac{162\pi x_c^5 \tilde{T}_c\tilde{P}_c}{4 \pi x_c^2\left(3x_c-\tilde{k}\right)^4}\left[3(-1+a)\left[2\tilde{k}^4+x_c\left(-23\tilde{k}^3+6 x_c(16\tilde{k}^2-27 x_c\tilde{k}+9 x_c)\right)\right]\right].\nonumber
\end{equation}

\end{widetext}
The pressure of the system remains constant during transition of phase from SBH to LBH when $t\,<\,0$. Hence, from Eq. (\ref{state}), we can write 
\begin{equation}\label{state2}
1+\mathcal{ U}t+\mathcal{ V}\epsilon_s-\mathcal{ W}\epsilon_{s}^{3}=1+\mathcal{ U}t+\mathcal{ V}\epsilon_l-\mathcal{ W}\epsilon_{l}^{3},
\end{equation}  where $\epsilon_{s}$ and $\epsilon_{l}\,$, respectively, are related to the radii of SBH and LBH.
To calculate the exponent $\beta$, the well known Maxwell's area law is used \cite{Nam:2019clw,Hyun:2019gfz}.
\begin{equation}\label{Maxwell}
\int_{\omega_s}^{\omega_l} \omega d\tilde{P} \, = \, 0.
\end{equation}
By using Eq. (\ref{state1}) and Eq. (\ref{omega}) in Eq. (\ref{Maxwell}), we get
\begin{equation}\label{state3}
 \frac{1}{2}\mathcal{V}\epsilon_l^2+\frac{3}{4}\mathcal{W} \epsilon_l^4=\frac{1}{2}\mathcal{V}t\epsilon_s^2+\frac{3}{4}\mathcal{W} \epsilon_s^4.
 \end{equation}
By solving Eqs. (\ref{state2}) and (\ref{state}), one gets
\begin{equation}
-\epsilon_s\,=\,\epsilon_l\,=\,\sqrt{\dfrac{-\mathcal{ V}t}{\mathcal{ W}}}.
\end{equation} 
Hence, the order parameter $\eta$ can be calculated as
\begin{equation}
\eta\,=\,\tilde{V_l}-\tilde{V_s}\,=\,\tilde{V}_c(\omega_l-\omega_s) \, \propto \, \sqrt{-t},
\end{equation} which means $\beta=1/2$. To calculate $\gamma$ we use the equation
\begin{equation}
\kappa_{\tilde{T}}\,=\,- \dfrac{1}{\tilde{V}}\,\dfrac{\partial\tilde{V}}{\partial\tilde{P}}\,\vert_{\,\tilde{T}}\,=\,\dfrac{4}{\tilde{P}\left(1+\omega\right)}\left(1-\dfrac{\tilde{k}}{2\,x_{c}^2}\right)\left(\dfrac{\partial \epsilon}{\partial p}\,\vert_{\,t}\right)\,\propto\,\dfrac{1}{\mathcal{ V}\,t},
\end{equation} which leads to $\gamma\,=\,1$. To compute $\delta$ we find the value of $\vert\,\tilde{P}\,-\,\tilde{P}_c\vert$ at $\tilde{T}\,=\,\tilde{T}_c$ or $t\,=\,0$, with the help of the equation
\begin{equation}
|\tilde{P}-\tilde{P}_c|_{\tilde{T}_c}\varpropto \epsilon^3\propto|\tilde{V}-\tilde{V}_c|^3,
\end{equation} which implies $\delta=3.$ The values of the critical exponents of the nonsingular-AdS black holes with clouds of strings background are similar to the Van der Walls fluid.   

\section{Conclusion}\label{conc} 
 The idea that the Schwarzschild black hole (point mass) may have atmospheres composed of clouds of strings because of the intense level of activity in string theory. The clouds of strings are the one-dimensional analogue of a dust cloud, have the same energy-momentum tensor as that of the global monopole, which could be effective on gravitational fields such as black holes. Further, it could describe a globular cluster with components of dark matter.   Further, it is firmly believed that the one-dimensional objects known as strings can describe the Universe; hence the study of black hole solution in clouds of strings background has a tremendous physical significance. Motivated by this and AdS/CFT correspondence, we have derived an exact nonsingular-AdS black hole solution endowed with clouds of strings background.  In turn, we have investigated the extended phase space of thermodynamics and studied the critical phenomena of the nonsingular-AdS black holes surrounded by clouds of strings background by treating the cosmological constant as a thermodynamical pressure. 

Interestingly, the thermodynamical quantities viz. temperature, pressure, specific heat and the Gibbs free energy are corrected except the entropy, which is not directly affected by the background clouds of strings, and then we probe the thermodynamic stability of the black hole. We find an analogy between the nonsingular-AdS black holes endowed with clouds of strings and Van der Waals fluid. Investigation of the  Hawking temperature revealed critical values of NED parameter $\tilde{k}$ and clouds of strings parameter ($a$) and the critical value of horizon radius $x_c$ where the temperature has local extrema  and also  divergence of specific heat thereby confirming the existing of two second order phase transition.  Interestingly, the heat capacity diverges at two critical radii $x_{c1}$ and $x_{c2}$, for $a<a_c$, respectively, at which the temperature has the local maximum and the local minimum values indicate that two second-order phase transitions exist in the canonical ensemble. The black hole unstable due to thermal fluctuation  when $x_{c1} < x < x_{c2}$ with negative specific heat and stable otherwise. The behaviour of Gibbs free energy revealed that the LBH with less Gibbs free energy are more stable than the SBH. We also analysed isobars on $\tilde{G}-\tilde{T}$ plane and found that when pressure $\tilde{P}$ is less than critical pressure $\tilde{P}_c$, the SBH underwent first-order phase transition (exhibiting swallow tail) to LBH at temperature $\tilde{T}_0$, and it was also found that an unstable IBH possible. Thus,  our model still exhibits SBH/LBH phase transition but at a larger horizon radius, and also two second-order phase transitions occur at a smaller horizon radius.   We found that the background clouds of strings impact the  $P-V$ critically and calculated the values of critical exponents to see that these values match that of Van der Walls fluid.     

Thus, we demonstrated that background clouds of strings profoundly influence the $P$--$V$ criticality and thermodynamic properties of black holes,   which may have several astrophysical consequences, for example, on wormholes and accretion onto black holes.  Some of the results presented here generalise previous discussions on nonsingular AdS black holes \cite{Kumar:2020cve} and Schwarzschild AdS black holes \cite{Dolan:2010} to a more general setting.  The possibility of further extensions of these results to higher curvature  gravity  is an interesting  problem for the future. 

\section{Acknowledgments} 
S.G.G. would like to thank SERB-DST for the ASEAN project IMRC/AISTDF/CRD/2018/000042.

\end{document}